\documentclass[aps,prb,groupedaddress,showpacs,twocolumn]{revtex4}

\usepackage{epsfig}
\usepackage{graphicx}
\usepackage{dcolumn}
\usepackage{amsmath}

\begin{document}
\bibliographystyle{apsrev}

\title{Quantitative study of molecular N$_{2}$ trapped in disordered GaN:O films}

\author{B.J. Ruck}
\email[e-mail: ]{b.ruck@irl.cri.nz}

\author{A. Koo}

\author{U.D. Lanke}

\author{F. Budde}

\author{S. Granville}

\author{H.J. Trodahl}

\affiliation{School of Chemical and Physical Sciences, Victoria
University of Wellington, P.O. Box 600, Wellington, New Zealand}

\altaffiliation{Alternative address: MacDiarmid Institute for Advanced Materials and Nanotechnology, Victoria
University of Wellington, P.O. Box 600, Wellington, New Zealand}

\author{A. Bittar}

\affiliation{Measurement Standards Laboratory, Industrial Research
Ltd., P.O. Box 31310, Wellington, New Zealand}

\author{J.B. Metson}

\affiliation{Department of Chemistry, Auckland University, Private Bag 92019, Auckland, New Zealand}

\author{V.J. Kennedy}

\author{A. Markwitz}

\affiliation{Institute of Geological and Nuclear Sciences, P.O. Box 30368, Lower Hutt, New Zealand}

\date{\today}

\begin{abstract}

The structure of disordered GaN:O films grown by ion-assisted
deposition is investigated using x-ray absorption near-edge
spectroscopy and Raman spectroscopy. It is found that between 4
and 21\,\% of the nitrogen in the films is in the form of
molecular N$_2$ that interacts only weakly with the surrounding
matrix. The anion to cation ratio in the GaN:O host remains close
to unity, and there is a close correlation between the N$_2$
fraction, the level of oxygen impurities, and the absence of
short-range order in the GaN:O matrix.

\end{abstract}

\pacs{68.55.-a, 78.66.Jg, 61.10.Ht}

\maketitle
\flushbottom

\section{Introduction}
\label{intro}

Determining the atomic scale structure and the resulting
electronic properties of complex material systems is an ongoing
challenge in physics research. An especially interesting material
with a complex structure is disordered gallium nitride (GaN). It
has been predicted that fully amorphous GaN has a well-defined
band gap relatively free of in-gap states, and may provide a
useful alternative to its crystalline
counterpart.\cite{Stumm_Drabold,Yu_Drabold} However, the amorphous
state is not in itself unique, and different fabrication
techniques give rise to a range of disordered structures.
Experimental studies have shown that the resulting optical and
electronic properties depend sensitively on the details of this
nano-structure, including the presence of nanocrystallites and
impurities.\cite{Bittar_Markwitz,Lanke_Trodahl,Koo_Trodahl1,Lanke_Bittar,Koo_Trodahl2,Metson_Bittar,Kang_Ingram,Kuball_Westwood,Chen_Kordesch,Miyazaki_Ohtsuka,Preschilla_Srinivasa,Yamada_Asami,Yang_Lee}

In this paper we report x-ray absorption near-edge spectroscopy
(XANES) and Raman spectroscopy results from disordered GaN films
containing varying levels of oxygen impurities (GaN:O). Our
principal observation is that between 4 and 21\,\% of the nitrogen
in these films is trapped as molecular N$_2$, interacting only
weakly with the surrounding solid matrix. We have previously shown
that the films grown by this technique range from nanocrystalline,
with crystallite size of order 3\,nm, to fully amorphous, with the
latter structure correlated with larger levels of oxygen
impurities.\cite{Lanke_Trodahl,Koo_Trodahl1} Here we observe also
a close correlation between the level of oxygen and the N$_2$
fraction, and we speculate on the nature of this three-way
relationship.

\section{Experimental Details}
\label{expt}

The thin films used in this study were grown by ion-assisted
deposition (IAD), where Ga atoms are deposited onto a substrate in
the presence of an energetic beam of nitrogen
ions.\cite{Bittar_Markwitz,Lanke_Trodahl,Koo_Trodahl1,Lanke_Bittar}
The ion energy was typically 500\,eV, with a total ion current
density of 0.2 -- 0.4\,mA/cm$^2$ and an overall growth rate of
around 0.5\,\AA/s. The resulting films were between 50 and
1000\,nm in thickness. Excellent adhesion was obtained onto a
range of substrates, including silicon, quartz glass, carbon
glass, mylar, and stainless steel. The substrate temperature
remained below 100$^{\circ}$C during growth.

Two separate vacuum systems were used for the film growth, with
different base pressures of $5\times10^{-9}$ and
$5\times10^{-6}$\,torr. Rutherford backscattering spectroscopy
(RBS) and nuclear reaction analysis (NRA) show that films grown in
the latter system contain significant amounts of oxygen, as
described below, while films grown with the lowest base pressure
have low oxygen content close to the measurement resolution of 1
atomic~\% (at.\%).\cite{Kennedy_Bittar}  Elastic recoil detection
(ERD) analysis found hydrogen fractions of 0.5 and 7\,at.\% in two
representative films containing 1 and 20\,at.\% oxygen,
respectively. For all samples the carbon impurity content was
below 1\,at.\%. The microstructure and degree of crystallinity of
the films was determined by x-ray diffraction (XRD), extended
x-ray absorption fine structure (EXAFS), and transmission electron
microscopy (TEM). All of the films described here are transparent
across the visible and show an absorption edge at around
3.5\,eV.\cite{Bittar_Markwitz,Koo_Trodahl1,Koo_Trodahl2}

Nitrogen K-edge XANES measurements were performed on the Canadian
SGM beamline at the Synchrotron Radiation Center, Madison,
Wisconsin. The monochromator has a maximum resolution of about
70\,meV. Data were taken in both the electron and fluorescent
yield modes (TEY and FLY), although we focus below on FLY data
where the penetration depth is about 70 -- 100\,nm and charging
effects are not significant.\cite{Katsikini_Akasaki} The Raman
data reported here were obtained at room temperature from samples
grown on silicon using a Jobin-Yvon LabRam spectrometer.
Excitation came from a 325\,nm HeCd laser with a focus spot size
of about 2\,$\mu$m, and a power of 1 -- 5\,mW.

\section{Results}
\label{results}

\begin{figure}
  \centering
  \includegraphics[width=7.0cm,keepaspectratio]{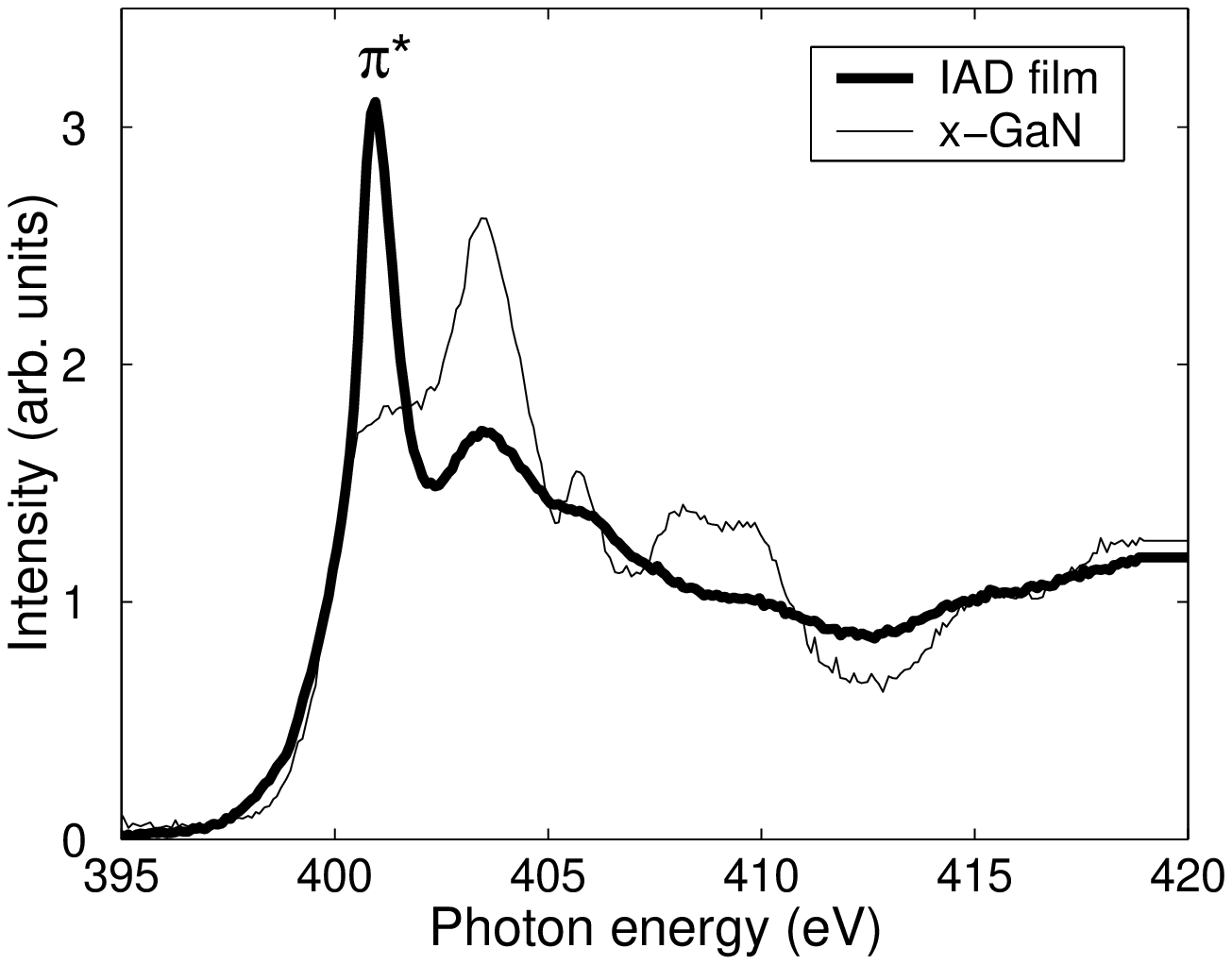}
  \includegraphics[width=7.0cm,keepaspectratio]{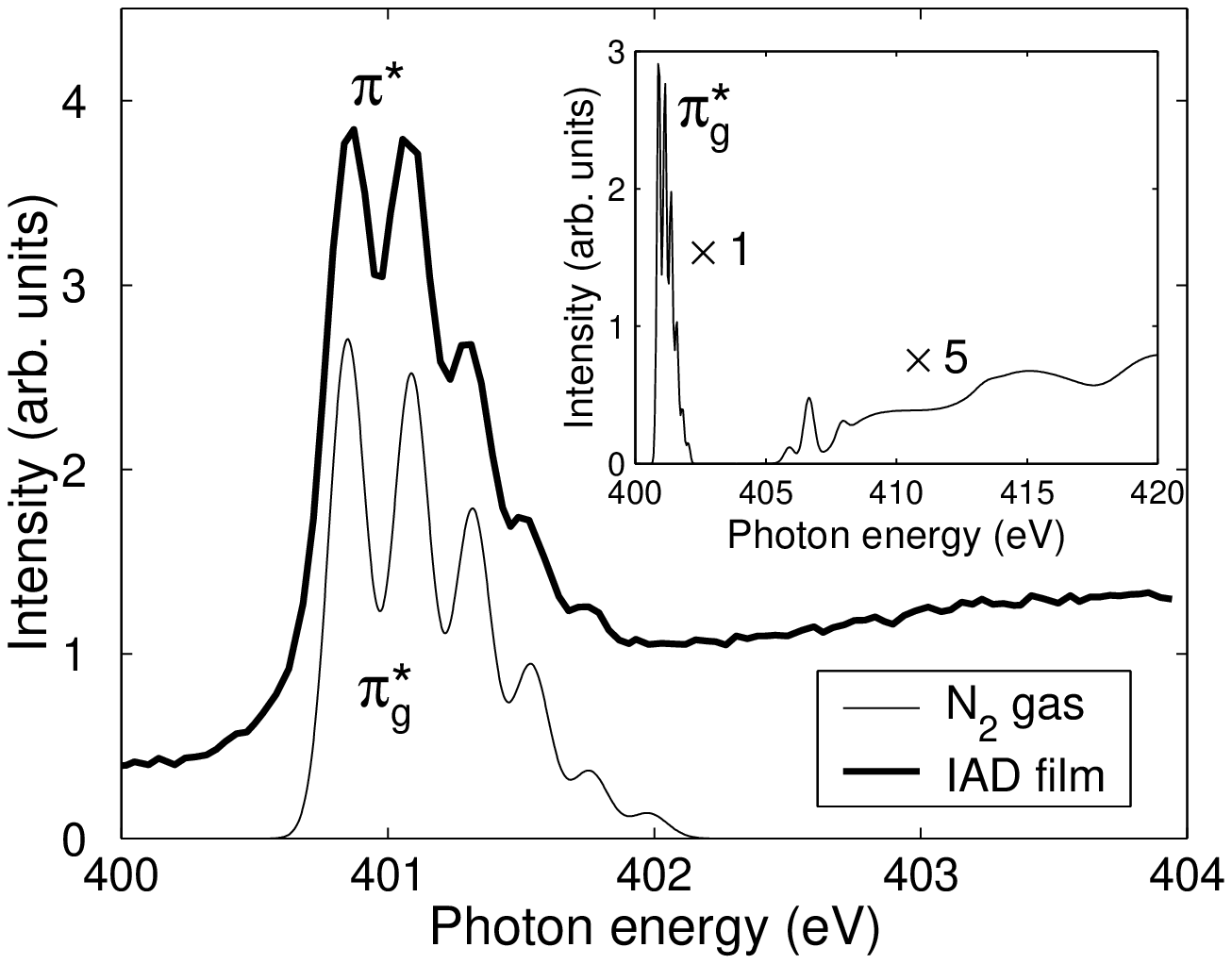}
  \includegraphics[width=7.0cm,keepaspectratio]{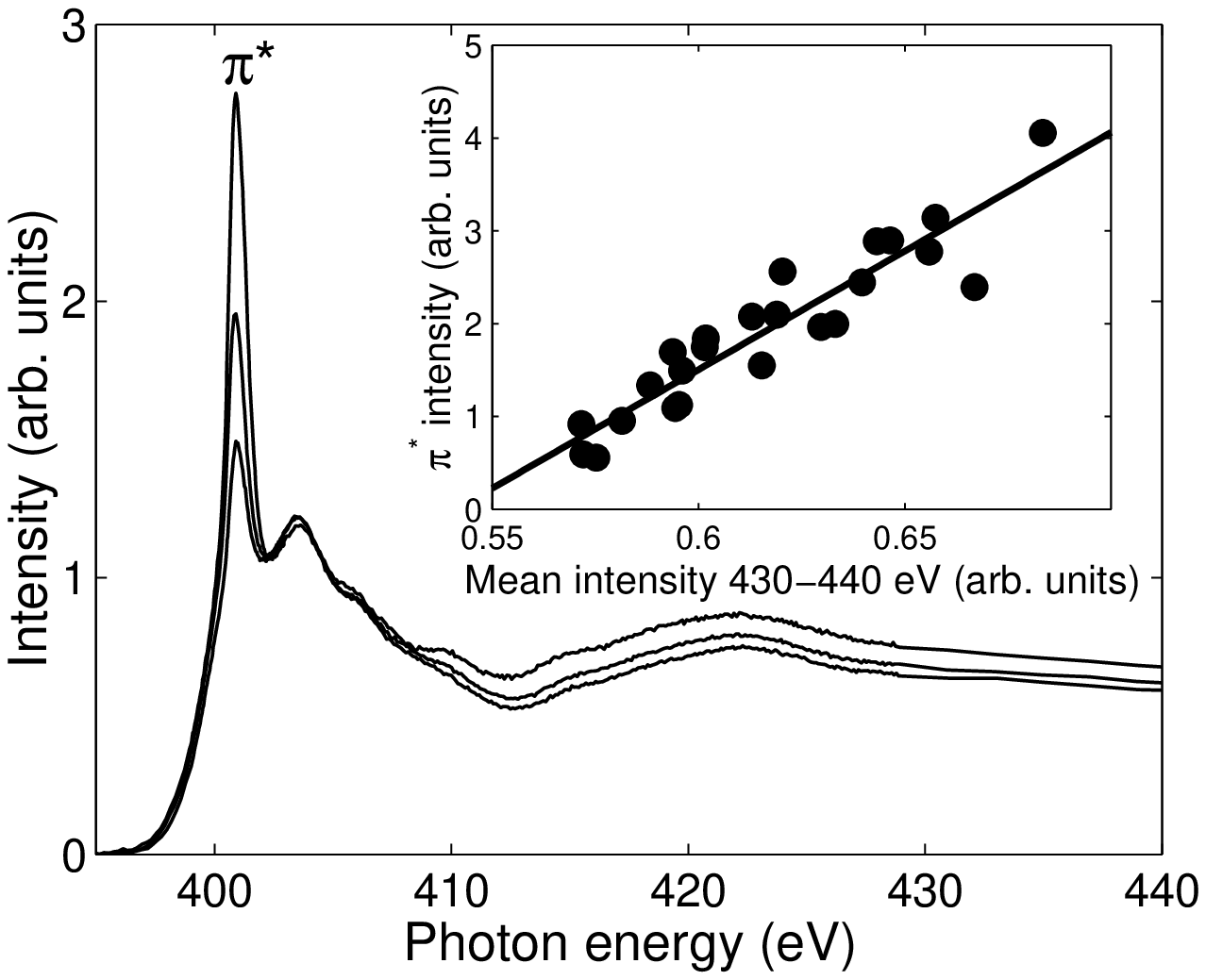}
\caption{(a)~XANES spectra from crystalline GaN and disordered
GaN:O. (b)~High-resolution $\pi^{*}$ and $\pi_{g}^{*}$ resonance
spectra from trapped and gaseous N$_2$, respectively. Inset:
Gaseous N$_2$ spectrum including contribution from transitions to
high energy states.\cite{Wenzel_Ertl} (c)~XANES spectra from
several disordered GaN:O films containing differing amounts of
trapped N$_2$. Inset: $\pi^{*}$ absorption versus average
absorption between 430 and 440\,eV for a series of samples.}
  \label{N2XANES}
\end{figure}

In Figure~\ref{N2XANES}(a) we compare the nitrogen K-edge XANES
spectrum from an IAD grown GaN:O sample containing about 8\,at.\%
oxygen with the spectrum from a single crystal c-axis oriented
wurtzite GaN film (x-GaN). The x-GaN spectrum has an absorption
onset at around 397\,eV followed by a series of peaks that can be
understood in terms of the $p$-projected calculated density of
states of wurtzite
GaN.\cite{Lambrecht_Wickenden,LawniczakJablonska_LilientalWeber}
The absorption onset of the IAD grown film also lies at about
397\,eV, but unlike the x-GaN sample, a large narrow peak centred
at 401\,eV dominates the spectrum. At higher energies the IAD
sample spectrum resembles a much-broadened version of the x-GaN
spectrum, as expected for these strongly disordered films.

A high-resolution scan over the 401\,eV peak reveals that it
actually consists of at least five narrow absorption lines, the
widths of which are limited by the monochromator resolution
[Fig.~\ref{N2XANES}(b)]. Also shown is the XANES spectrum of
gaseous molecular nitrogen measured on the same beam-line, showing
the vibrational splitting of the $2p\ \pi_g^{*}$ resonance (the
energy scale was calibrated using the known location of this
peak).\cite{Yates_Bancroft} The similarity in peak positions and
relative intensities allows us to attribute the 401\,eV peak to
the $\pi^{*}$ resonance of molecular nitrogen trapped within the
films.\cite{Metson_Bittar} The inset to Fig.~\ref{N2XANES}(b)
shows the spectrum of gaseous molecular nitrogen expanded to
include the contribution from transitions to high energy Rydberg
and multiple-electron excited states that begin at about 406\,eV
(N$_2$ absorption data above 406\,eV taken from
Ref.~\onlinecite{Wenzel_Ertl}).

The fraction $F_{\mathrm{N}_2}$ of the nitrogen held in molecular
form varies between samples, as illustrated in
Fig.~\ref{N2XANES}(c) by the variation in the relative intensity
of the $\pi^{*}$ peak (the normalization procedure is described
below). Interestingly, the films with the greatest N$_2$ fraction
are those that show the most broadening in the underlying GaN:O
absorption spectrum. Structural characterization by XRD, TEM, and
EXAFS has indicated that these films are fully amorphous, whereas
the films with lower N$_2$ levels contain crystallites of
approximate size 3\,nm.\cite{Koo_Trodahl1} Furthermore, RBS and
NRA measurements show that the amorphous films also contain high
levels of oxygen, typically greater than 15 at.\,\%, and it may be
that a significant oxygen level is essential to stabilize the
amorphous structure.

An estimate of the value of $F_{\mathrm{N}_2}$ for each sample can
be obtained in the following fashion. Firstly, we note that the
N$_2$ makes no contribution to the absorption over the energy
range 403 -- 406\,eV, so the entire contribution in this window is
from the GaN:O matrix. The spectra in Fig.~\ref{N2XANES}(c) have
been normalized over this range, resulting in an offset that
develops above 406\,eV representing the extra absorption due to
transitions to high energy states in the trapped N$_2$. As
expected, the offset is proportional to the intensity of the
$\pi^*$ peak, and thus to $F_{\mathrm{N}_2}$.

At photon energies far above the edge the absorption cross section
becomes independent of the bonding environment of the absorbing
nitrogen atom. (EXAFS oscillations are of insignificant amplitude
here). Therefore, we examine the energy range 430 -- 440\,eV,
where there is no evidence of structure in the absorption, and
plot the average amplitude in this range against the fitted
$\pi^*$ peak amplitude for each sample (see Fig.~\ref{N2XANES}(c)
inset). Extrapolating these data shows that the absorption between
430 and 440\,eV would be about 0.54 (on this scale) in the absence
of any N$_2$.  The actual absorption varies from 0.57 to 0.69, so
the N$_2$ contributes between 4 and 21\,\% of the total
absorption. Thus we estimate that $F_{\mathrm{N}_2}$ varies from
0.04 to 0.21.

It should be noted that the magnitude of the contribution from the
GaN:O matrix between 403 and 406\,eV depends on the density of
$p$-projected electronic states in this range. This in turn may
depend on the details of the microstructure, such as the degree of
crystallinity. For this reason we chose not to use the x-GaN
spectrum to estimate the absorption when $F_{\mathrm{N}_2}=0$, and
instead used only the disordered samples where, apart from the
N$_2$ contribution, there are only subtle variations amongst the
spectra. Although the normalization, and hence the value of
$F_{\mathrm{N}_2}$, is only approximate, our finding that up to
21\,\% of the nitrogen is in the form of N$_2$ is consistent with
independent estimates described below. We also note that the
distinct resonant peaks above 406\,eV in the gaseous N$_2$
spectrum (Fig.~\ref{N2XANES}(b) inset) are not apparent in the
offset region of the IAD GaN:O spectra. This is consistent with
results from adsorbed films of N$_2$, where interaction with the
substrate usually broadens the resonances into a smooth
continuum.\cite{Wenzel_Ertl,Ruckman_Strongin}

\begin{figure}
  \centering
  \includegraphics[width=8.0cm,keepaspectratio]{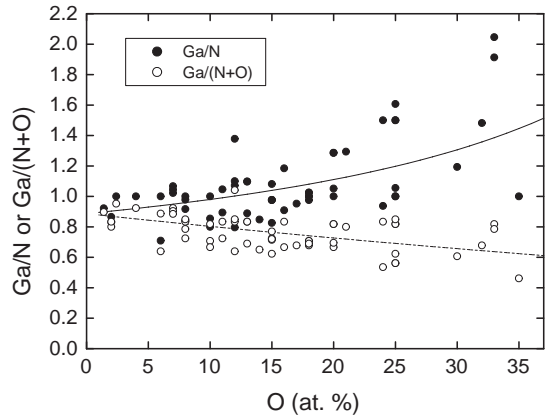}
\caption{Measured atomic ratios of gallium to nitrogen ($\bullet$)
and gallium to nitrogen plus oxygen ($\circ$), plotted versus
oxygen atomic percentage. The solid and dashed lines show the
expected ratios calculated using the N$_2$ atomic fraction
determined from XANES data.}
  \label{RBSvsO}
\end{figure}

The effect of oxygen incorporation on the stoichiometry,
determined by RBS and NRA, is plotted in Figure~\ref{RBSvsO}. The
Ga to N ratio tends to increase above unity as the oxygen
concentration increases. We attribute this to the substitution of
oxygen for nitrogen in the GaN matrix, so that the composition is
roughly GaO$_x$N$_{1-x}$. There are relatively few reports in the
literature on the properties of deliberately grown
GaO$_x$N$_{1-x}$, but it is known that high levels of oxygen
($x\approx 0.3$) can be substituted into GaN without substantially
altering the lattice structure.\cite{Aleksandrov_Zykov} We have
found no evidence for the formation of a separate gallium oxide
phase in TEM, XRD, EXAFS, or Raman measurements.

It was noted above that $F_{\mathrm{N}_2}$ increases with oxygen
content, with the minimum value 0.04 corresponding to near oxygen
free samples and the maximum value 0.21 being obtained when the
oxygen content $x=0.35$. Assuming a simple linear dependence we
write $F_{\mathrm{N}_2}=0.04+0.49x$. The solid and dashed lines in
Fig.~\ref{RBSvsO} show, respectively, the variation of the gallium
to nitrogen and gallium to total cation (oxygen plus nitrogen)
ratios expected given this form for $F_{\mathrm{N}_2}$. The trend
in both ratios is captured very well with no free parameters.

To the best of our knowledge there are no specific predictions in
the literature regarding the energetics of molecular nitrogen
defects in GaN.\cite{VanDeWalle_Neugebauer} Given our growth
technique, and the fact that all of the films contain some N$_2$,
it seems likely that the ion beam simply buries some N$_2$ during
growth. Indeed, TRIM calculations show that the penetration depth
of the 500\,eV N$_2^+$ ions is a few atomic layers. The cause of
the correlation between N$_2$ and oxygen content is less obvious.
It is known that interstitial N$_2$ is formed in TiN and CrN films
when N atoms are displaced during surface
oxidation.\cite{Esaka_Kikuchi} A similar oxidation process, taking
place as the growth proceeds, may add to the N$_2$ level in our
films. Alternatively, it may simply be that the more disordered
structure associated with the high oxygen content films is able to
trap more N$_2$. In any case, we stress that the GaN:O matrix is
stoichiometric in the sense that the anion to cation ratio,
excluding the N$_2$, is approximately one in all films. This is in
contrast to GaN amorphised by heavy-ion bombardment, in which
N$_2$ bubbles form from displaced nitrogen, leaving behind a
Ga-rich lattice.\cite{Kucheyev_Li2}

\begin{figure}
  \centering
  \includegraphics[width=8.0cm,keepaspectratio]{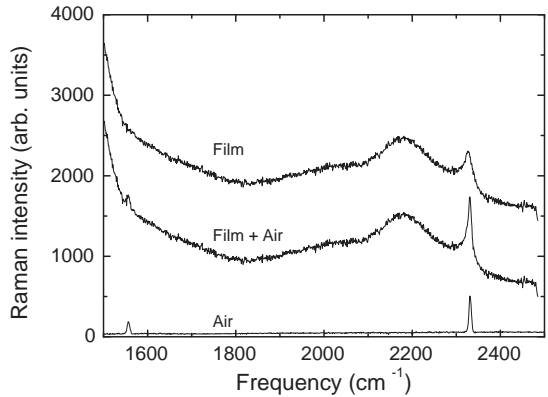}
\caption{Raman spectrum from a 130\,nm thick film with negligible
oxygen content, showing the molecular N$_2$ vibration at
2360\,cm$^{-1}$, as well as the broad third order GaN mode at
2200\,cm$^{-1}$. The curve labelled \textsf{Film} has had the
contribution from atmospheric N$_{2}$ removed to give only the
spectrum of the N$_2$ trapped in the film.}
  \label{RamanN2}
\end{figure}

A portion of the Raman spectrum of a low oxygen
($\approx1$\,at.\%) IAD grown film, showing the broad third order
GaN mode at 2200\,cm$^{-1}$, is shown in Figure~\ref{RamanN2}. For
comparison, the spectrum from a single crystal GaN film may be
found in Ref.~\onlinecite{Behr_Akasaki}. Also apparent is a narrow
N$_2$ vibrational mode at 2360\,cm$^{-1}$, which represents a
combination of scattering from N$_2$ in the film and free N$_2$ in
the air between the sample and the microscope objective. We
eliminate the latter by subtracting the signal obtained from a
bare silicon substrate, where both spectra are first normalized
using the molecular oxygen mode at 1556\,cm$^{-1}$ that originates
solely from the air. The resulting spectrum, representing only
N$_2$ in the film, is also shown in Fig.~\ref{RamanN2}. Note that
Raman probes the vibrational splitting of the electronic ground
state of N$_2$, whereas XANES probes the same vibrationally split
anti-bonding $\pi^{*}$ state. The Raman peak at 2360\,cm$^{-1}$,
which corresponds to about 0.29\,eV, is therefore at higher energy
than the splitting between the lowest vibrational levels of N$_2$
observed in XANES (0.24\,eV).

The laser spot diameter was about $2\,\mu\mathrm{m}$, so for the
confocal Raman microscope the air signal comes from a volume just
above the sample of roughly  $8\,\mu\mathrm{m}^{3}$, containing
about $1.6\times10^{8}$ N$_2$ molecules. The GaN:O absorption
coefficient at 325\,nm ($10^{5}$\,cm$^{-1}$) allows the beam to
probe the top $50\,$nm of the
film.\cite{Koo_Trodahl1,Koo_Trodahl2} As the trapped N$_2$ peak
has roughly 4 times the area of the peak from N$_2$ in the air,
the probed volume of the film contains approximately
$6.4\times10^{8}$ N$_2$ molecules, corresponding to a density of
$3.2\times10^{21}$\,N$_{2}\,\mathrm{cm^{-3}}$. Using the atomic
density of crystalline GaN
($8.8\times10^{22}\,\mathrm{at.\,cm^{-3}}$) to estimate the
density of our films we find that about 3\,\% of the nitrogen is
in the form of N$_2$. This is in excellent agreement with our
estimate of 4\,\% based on the XANES results for this sample.

The width of the trapped N$_2$ peak is about 30\,cm$^{-1}$,
considerably broader than the corresponding value of 10\,cm$^{-1}$
for free N$_2$, and the broadening is most pronounced on the low
energy side. The peak is also shifted to lower energy by about
4\,cm$^{-1}$. The fractional peak width $\Delta\omega/\omega$ is
only about 1\,\%, implying that the nitrogen molecules experience
a weak, but non-zero interaction with the host matrix. The
presence of this interaction suggests that the N$_2$ probably
resides in interstitial sites in the films rather than within
voids or cracks.

It is interesting to question whether subsequent film treatments,
such as thermal annealing, can influence the trapped N$_2$. To
this end we have annealed a nanocrystalline sample containing
about 10\,at.\% oxygen at 700$^{\circ}$C for 30 minutes under
flowing nitrogen. After annealing the XANES features related to
the GaN:O matrix become somewhat sharper, and the $\pi^{*}$ peak
decreases in intensity by approximately 20\% indicating some loss
of N$_2$ from the film.

In conclusion, we have used a range of experimental probes to
characterize a series of highly disordered GaN:O films. A detailed
analysis of the data has demonstrated quantitatively the presence
of molecular N$_2$ in the films. Interaction with the host matrix
leads to only a small broadening of the N$_2$ vibrational
frequency. Both the atomic-scale structure of the GaN:O and the
amount of N$_2$ depend on the growth conditions, especially the
level of oxygen impurities.  These results highlight the
sensitivity of the properties of such highly disordered materials
to small variations in the preparation conditions.

\begin{acknowledgments}
The authors are grateful to Yong Feng Hu and Astrid J\"{u}rgenson
of the Canadian Synchrotron Source for experimental assistance. We
acknowledge the New Zealand Foundation for Research, Science and
Technology for financial assistance through its New Economy
Research Fund, and through doctoral (AK) and postdoctoral (BJR)
fellowships.
\end{acknowledgments}

\clearpage

\end{document}